# Quantum Dense Coding Exploiting Bright EPR Beam

Xiaoying Li, Qing Pan, Jietai Jing, Jing Zhang, Changde Xie & Kunchi Peng

*The Key Laboratory for Quantum Optics, Ministry of Education, China; The Institute of Opto-electronics, Shanxi University, Taiyuan, 030006, P. R. China*

**Highly efficient quantum dense coding for continuous variables has been experimentally accomplished by means of exploiting bright EPR beam with anticorrelation of amplitude quadratures and correlation of phase quadratures, which is generated from a nondegenerate optical parametric amplifier operating in the state of deamplification. Two bits of classical information are encoded on two quadratures of a half of bright EPR beam at the sender Alice and transmitted to the receiver Bob via one qubit of the shared quantum state after encoding. The amplitude and phase signals are simultaneously decoded with the other half of EPR beam by the direct measurement of the Bell-state at Bob. The signal to noise ratios of the simultaneously measured amplitude and phase signals are improved 5.4dB and 4.8dB with respect to that of the shot noise limit respectively. A high degree of immunity to unauthorized eavesdropping of the presented quantum communication scheme is experimentally demonstrated.**

  Quantum teleportation and quantum dense coding are two typical examples to exploit nonlocal quantum correlation of entangled states in quantum information to perform otherwise impossible tasks. Quantum teleportation is the disembodied transport of an unknown quantum state from one place to another[1]. Quantum dense coding provide a method by which two bits of classical information can be transmitted by sending one qubit of quantum information[2]. Discrete and continuous variable teleportations have been performed experimentally for single-photon polarization states[3,4,5] and coherent state of electromagnetic field modes[6], respectively. The dense coding, originally proposed[1], successively discussed and experimentally demonstrated, are mostly concentrated within the setting of discrete quantum variables[7,8,9]. Later, the schemes realizing highly efficient dense coding for continuous variables are theoretically proposed, in that the two-mode squeezed-state entanglement are exhibited to achieve unconditional signal transmission[10,11,12]. So far, the experimental demonstration on the quantum dense coding for continuous variables has not been presented to our knowledge. Our motivation to realize experimentally the dense coding for continuous quantum variables is not only for that its information capacity always beats signal-mode coherent-state communication and surpasses signal-mode squeezed state communication if the average photon numbers $\bar{n} > 1$[10] also for its high degree of immunity to unauthorized eavesdropping, since the signal-to-noise ratios in information channel are very small if without available other half of EPR beam to decode the signals, and furthermore, any attempt to extract information will reveal itself by a decrease in the signal-to-noise ratio and an increase of quantum fluctuation. Therefore the dense coding for continuous variables with enough robust quantum correlation has potentially application for future quantum communication and the developing quantum information processing[13,14].

As well known, a continuous nondegenerate optical parametric amplifier (NOPA) operating in the state of amplification produces EPR beam with the correlation of amplitude quadratures and anticorrelation of phase quadratures, i.e. the variances of the difference of amplitude quadratures $\langle \delta(X_1 - X_2)^2 \rangle$ and the sum of phase quadratures $\langle \delta(Y_1 + Y_2)^2 \rangle$ are both smaller than the shot noise limit (SNL) defined by the vacuum fluctuation[15,16], and for NOPA operating at deamplification the quantum correlations of quadrature components between the signal and idler modes have inverse signs, i.e. $\langle \delta(X_1 + X_2)^2 \rangle$ <SNL and $\langle \delta(Y_1 - Y_2)^2 \rangle$ <SNL[17]. It has been theoretically proved in our previous publication[11] that the bright EPR beam having nonzero average intensity with the anticorrelation of amplitude quadratures[$\langle \delta(X_1 + X_2)^2 \rangle \to 0$] and the correlation of phase quadrature[$\langle \delta(Y_1 - Y_2)^2 \rangle \to 0$] can be used for realizing quantum teleportation and dense coding of continuous variables by means of the direct detection for photo-currents and two radio frequency (rf) power splitters. Based on the proposed scheme, we performed the first dense coding experiment for continuous variables in which simultaneous measurements of phase and amplitude signals, with a sensitivity beyond that of the SNL, were achieved.

The experimental diagram is shown in Fig.1. The entangled EPR beam is generated from a NOPA consisting of an α–cut type II KTP (KTiOPO$_4$, Potassium Titanyl Phosphate) crystal (10mm long), the front face of which is coated to be used as the input coupler( the transmission>95% at 540nm wavelength and 0.5% at 1080nm) and the other face is coated with the dual-band antireflection at both 540nm and 1080nm, as well as a concave mirror of 50mm-curvature radius, which is used as the output coupler of EPR beam at 1080nm (the transmission of 5% at 1080 and high reflectivity at 540nm). The output coupler is mounted on a piezoelectric transducer to lock actively the cavity length on resonance with the injected seed wave at 1080nm using the FM sideband technique. By fine tuning the crystal temperature the birefringence between signal and idler waves in KTP is compensated and the simultaneous resonance in the cavity is reached. The process of adjusting temperature to meet double resonance can be monitored with an oscilloscope during scanning the length of cavity. Once the double resonance is completed the NOPA is locked on the frequency of the injected seed wave[16]. The measured finesse, the free spectral range, and the line-width of the parametric oscillator are 110, 2.8G, and 26 MHz, respectively. The pump source of NOPA is a home-made all-solid-state intracavity frequency-doubled and frequency-stabilized CW ring Nd:YAP (Nd:YAlO$_3$, Yttrium-Aluminum-Perovskite) laser, the output second-harmonic wave at 540nm and the leaking fundamental wave at 1080nm from which serve as the pump light and the seed wave respectively. The laser-diode pumped all-solid-state laser and the semi-monolithic F-P configuration of the parametric cavity ensure the stability of system. So that the frequency and phase of light waves can be well-locked during proceeding the dense coding experiment. Locking the relative phase between the pump laser and injected seed wave of NOPA to (2n+1)π, where n is integer, to enforce the NOPA operating at deamplification, the entangled EPR beam with the quantum anticorrelation of amplitude quadratures and the correlation of phase quadratures was generated. The two halves of EPR beam are just the signal and idler modes of the light field produced from type II parametric down conversion, thus they have orthogonal polarizations originally. The correlations measured by self-homodyne detector[11,18] between the quardrature-phase amplitudes of the two halves of EPR beam are show in Fig2. Both variances of $\langle\boldsymbol{\delta}(X_1+X_2)^2\rangle$ (red line) and $\langle\boldsymbol{\delta}(Y_1-Y_2)^2\rangle$ (blue line) measured directly are ~4dB below that of the SNL (considering the electronics noise that is 8dB below the SNL, the actual fluctuation should be ~5.4dB below that of the SNL). The bright EPR beam of ~70μW was obtained at the following operation parameters of NOPA: the pump power 150mW just below the power of the oscillation threshold of 175mW and the polarization of that was along the b axis of the KTP crystal. The power of the injected seed wave was 10mW before entering the input coupler of the cavity and that was polarized at $45^0$ relative to the b axis.

The two halves ( 1 and 2 ) of EPR entangled beam were distributed to the sender Alice and the receiver Bob, respectively. At Alice the classical amplitude and phase signals were modulated on the one half of EPR beam (beam 1), which led to a displacement signal sent via the quantum channel. Due to that each half of EPR beam had huge noise individually, for perfect EPR entanglement $\langle\delta(X_{1(2)})^2\rangle\to\infty$ and $\langle\delta(Y_{1(2)})^2\rangle\to\infty$, the signal to noise ratios in the beam 1 for both amplitude and phase signals tended to zero, so no one other than Bob can gain any signal information from the modulated beam under ideal condition. At Bob, the signals were decoded with the aid of the other half of the EPR beam (beam 2), which was combined with the modulated first half of EPR beam (beam 1) on a 50% beam-splitter, in our experiment which consisted of two polarized-beamsplitters and a half-wave-plate[19], and before combination, a π/2 phase shift was imposed between them according to the requirement of direct measurement of Bell-State[11]. The bright outcomes from two ports of the beamsplitter were directly detected by a pair of photodiodes $D_1$ and $D_2$(ETX500 InGaAs), and then each photocurrent of $D_1$ and $D_2$ was divide into two parts through the power splitters. The sum and difference of the divided photocurrent were nothing else but the transmitted amplitude and phase signals from Alice and Bob, which are expressed by[11]

$$i_+(\Omega) = \frac{1}{\sqrt{2}}\{[X_1(\Omega) + X_2(\Omega)] + X_s(\Omega)\} \quad (1)$$

$$i_-(\Omega) = \frac{1}{\sqrt{2}}\{[Y_1(\Omega) - Y_2(\Omega)] + Y_s(\Omega)\} \quad (2)$$

where $X_s(\Omega)$ and $Y_s(\Omega)$ are the modulated amplitude and phase signals on the first half of EPR beam at the sender Alice. With Perfect EPR entangled beam $\langle \delta(X_1 + X_2)^2 \rangle \to 0$ and $\langle \delta(Y_1 - Y_2)^2 \rangle \to 0$, Eqs.(1) and (2) are simplified:

$$i_+(\Omega) = \frac{1}{\sqrt{2}}\{X_s(\Omega)\} \quad (3)$$

$$i_-(\Omega) = \frac{1}{\sqrt{2}}\{Y_s(\Omega)\} \quad (4)$$

It means that under ideal conditions, the signals $X_s(\Omega)$ and $Y_s(\Omega)$ encoded on the amplitude quadrature and phase quadrature of beam 1 are simultaneously retrieved at the receiver Bob without any error. In general, the both encoded signals will be recovered with a sensitivity beyond that of the SNL when the beam 1 and beam 2 are quantum entangled. In fact, the sum of the amplitude quadratures between the two halves of EPR beam commutes with the difference of its phase quadrature, therefore the detected variances in them can be below that of the SNL simultaneously, and not violate the uncertainty principle[20]. Fig.3 shows the directly measured amplitude(red line) and phase (blue line) signals at Bob, which are the signals of 2MHz modulated on the first half of EPR beam (beam 1) by the amplitude and phase modulators at Alice. It can be seen that the original signals are retrieved with the high signal to noise ratio of ~4dB and ~3.6dB beyond that of the SNL under the help of the other half of EPR beam (beam 2)(accounting for the electronics noise of ~8dB below the SNL, the actual value should be ~5.4dB and ~4.8dB respectively) . Compared with the previously completed sub-shot-noise-limit optical measurements and quantum non-demolition measurements for a signal, in which the squeezed state light have been applied[20-26], our experiments have achieved the simultaneous measurements of two signals modulated on the amplitude and phase quardratures respectively with the precision beyond that of the SNL by means of exploiting the EPR entanglement. Although when sending and modulating two coherent beams a factor of two in channel capacity may also be gained, the signal-to-noise ratios of measurements are not able to breakthrough the SNL.

As well-discussed in Ref[27], in which a signal is transmitted via two quantum channels of EPR pair, in our scheme the individual signal channel has a high degree of immunity to unauthorized interception since very low signal-to-noise ratios. Fig.4. demonstrates the security of the signal channel against eavesdropping, where the red line is the fluctuation spectrum of the first half of EPR beam 1 with the modulated signals measured individually at Bob while the other half of EPR beam is not applied, which is ~4dB (considering the electronics noise that is 5dB below the SNL, the actual fluctuation should be ~5.4dB)above that of the corresponding SNL (black line). It is obvious that no any signal can be extracted since the signals are submerged in the large noise background totally.

Based on our previously proposed theoretical protocol[11], the first dense coding for continuous variables has been experimentally realized by exploiting bright EPR beam from NOPA operating at deamplification and the direct measurement of Bell-state. The mature technique for producing EPR beam from the parametric down conversion and the simplicity of direct measurement, as well the large coding capacity and the security against eavesdropping provided by quantum entanglement, make the presented scheme valuable for the experimental realization of the proposed theoretical protocols of quantum information processing and quantum communication, such as quantum teleportation[11], quantum cryptography[28], quantum error correction[13,14], quantum purification[29] and quantum cloning[30] etc.. The product of the conditional variances of the EPR beam obtained in our experiment is $\langle\Delta(X_1+X_2)^2\rangle\langle\Delta(Y_1-Y_2)^2\rangle = 0.62$. Assuming the EPR beam to be used for quantum teleportation on ideal conditions (gain=1, perfect transmission and detection efficiency), the yielded fidelity would be F=0.71(accounting for the electronics noise, the values should be 0.332 and 0.78 respectively).

Reference

1. Bennett, C. H. et al. Teleporting an unknown quantum state via dual classical and Einstein-Podolsky-Rosen channels. *Phys. Rev. Lett.* **70**, 1895-1899 (1993)

2. Bennett, C. H. & Wiesner, S. J. Communication via one- and two-particle operators on Einstein-Podolsky-Rosen states. *Phys. Rev. Lett.* **69**, 2881-2884( 1992)

3. Bouwmeesteret, D. et al. Experimental quantum teleportation. *Nature* (London) **390**, 575-579 (1997)

4. Boschi, D., Branca, S., De Martini, F., Hardy, L. & Popescu, S. Experimental realization of teleporting an unknown quantum state via dual classical and Einstein-Podolsky-Rosen channels. *Phys. Rev. Lett.* **80** 1121-1124 (1998)

5. Kim, Y. H., Sergei P., & Shih, Y. H.. Quantum teleportation of a polarization state with a complete Bell state measurement. *Phys. Rev. Lett.* **86**, 1370-1373 (2001)

6. Furusawa, A. et al. Unconditional quantum teleportation. *Science* **282**, 706-709 (1998)

7. Barenco, A. & Ekert, A. K. Dense coding based on quantum entanglement. *J. Mod. Opt.* **42**, 1253-1259 (1995)

8. Shimizu, K., Imoto, N. & Mukai, T. Dense coding in photonics quantum communication with enhanced information capacity. *Phys. Rev. A* **59,** 1092-1097 (1999)

9. Mattle, K., Weinfurter, H., Kwiat, P.G., and Zeilinger, A. Dense Coding in experimental quantum communication. *Phys. Rev. Lett.* **76**, 4656-4659 (1996)

10. Braunstrin, S. L. & Kimble, H. J. Dense coding for continuous variables. *Phys. Rev. A* **61**, 042302 (2000)

11. Zhang, J. & Peng, K. C. Quantum teleportation and dense coding by means of bright amplitude-squeezed light and direct measurement of a Bell state. *Phys. Rev. A* **62**, 064302 (2000)

12. Ban, M. Quantum dense coding via a two-mode squeezed-vacuum state. *J. opt. B: Quantum Semiclass. Opt.* **1** L9-L11(1999)

13. Braunstein, S. L. Error correction for continuous quantum variables. *Phys. Rev. lett.* **80**, 4084-4087 (1998)

14. Braunstein, S. L. Quantum error correction for communication with linear optics. *Nature* (London) **394**. 47-49 (1998)

15. Ou, Z. Y., Pereira, S. F., Kimble, H. J. & Peng, K. C. Realization of the Einstein-Podolsky-Rosen paradox for continuous variables. *Phys. Rev. Lett.* **68**, 3663-3666 (1992)

16. Zhang, Y. et al. Experimental generation of bright EPR beams from narrowband nondegenerate optical parametric amplifier. *Phys. Rev. A* **62**, 023813 (2000)

17. Zhang, Y., Su, H., Xie, C. D., & Peng, K. C. Quantum variances and squeezing of output field from NOPA. *Phys. Lett. A* **259**, 171-177 (1999)

18. Schneider, K., Bruckmeier, R., HanSen, H., Schiller,S. & Mlynek,J. Bright squeezed-light generation by a continuous-wave sememonolithic parametric amplifier, *Opt. lett.* **21**, 1396-1398 (1996)

19. Peng, K. C. et al. Generation of two-mode quardrature-phase squeezing and intensity difference squeezing from a cw-NOPO. *Appl. Phys. B* **66**, 755-758 (1998)

20. Ou, Z. Y., Pereira , S. F. & Kimble, H. J. Realization of the Einstein-Podolsky-Rosen paradox for continuous variables in nondegenerate parametric amplification. *Appl. Phys. B: photophys. Laser Cham* **55**, 265-278 (1992)

21. Xiao, M., Wu, L. A. & Kimble, H. J. Precision measurement beyond the shot-noise limit. *Phys. Rev. Lett.* **59**, 278-281 (1987)

22. Grangier, P., Slusher, R. E., Yurke, B. & La Porta, A. Squeezed-Light-Enhanced polarization interferometer. *Phys. Rev. Lett.* **59**, 2153-2156 (1987)

23. For a review, see Grangier, P., Levenson, J. A. & Poizat, J-P .Quantum non-demolition measurements in optics. *Nature* (London) **396**, 537 - 542 (1998)

24. Mertz, J., Debuisschert, T., Heidman, A., Fabre, C. & Giacobino, E. Improvement in the observed intensity correlation of OPO twin beams. *Opt. Lett.* 16, 1234-1236 (1991)

25. Ribeiro, P. H., Schwob, C., Matre, A. & Fabre, C. Sub-shot-noise high-sensitivity spectroscopy with optical parametric oscillator twin beams. *Opt. Lett*. **22**, 1893-1895 (1997)

26. Wang, H. et al. Experimental realization of a quantum measurement for intensity difference fluctuation using a beam splitter. *Phys. Rev. Lett.* **82**, 1414-1417 (1999)

27. Pereira, S. F., Ou, Z. Y. & Kimble, H. J. Quantum communication with correlated nonclassical states. *Phys. Rev. A.* **62**, 042311 (2000)

28. Ralph, T. C., Quantum cryptography with squeezed state. *Phys, Rev. A* **61**, 022309 (2000)

29. Parker, S., Bose, S., & Plenio, M. B., Entanglement quantification and purification in continuous-variable systems. *Phys, Rev. A* **62**, 052318 (2000)

30. Cref, N. J., Ipe, A., & Rottenberg, X. Cloning of Continuous Variables. *Phys, Rev. Lett*. **85**, 1754-1757 (2000)

Acknowledgement: The work is supported by the National Nature Science Foundation(No.69837010 and No. 19974021)

**Correspondence and requests for materials should be addressed to K.C. Peng (e-mail: kcpeng@sxu.edu.cn).).**

Fig.1 Schematic of the experimental apparatus for dense coding for Continuous variables. Two bits of classical information Xs and Ys are encoded on the amplitude and phase quadratures of a half of EPR beam (beam 1) at Alice, then are decoded by the other half of EPR beam (beam 2) at Bob.

Fig.2 Spectral densities of photocurrent fluctuations $<\delta(X_1+X_2)^2>$(red line) and $<\delta(Y_1+Y_2)^2>$(blue line ), SNL —the Shot Noise Limit (black line). Acquisition parameters: radio frequency (rf) $\Omega/2\pi$=2MHz, resolution bandwidth $\Delta\Omega/2\pi$=30KHz, Video bandwidth 0.1KHz, the electronics noise is 8dB below the SNL.

Fig.3 Measured amplitude (red line) and phase signal (blue line) at Bob, when EPR beam 1 is phase and amplitude modulated at 2MHz at Alice. SNL —the Shot Noise Limit (black line). Acquisition parameter: measured frequency range 1.0MHz-3.0MHz, resolution bandwidth 30KHz, video bandwidth 0.1KHz, the electronics noise is 8dB below the SNL.

Fig.4 Spectral density of photocurrent fluctuations of EPR beam 1 with the modulation signals (red line), the modulated signals are submerged in the noise background. SNL —the Shot Noise Limit (black line) Acquisition parameter: measured frequency range 1.0MHz-3.0MHz, resolution bandwidth 30KHz, vides bandwidth 0.1KHz, the electronics noise is 5dB below the SNL.

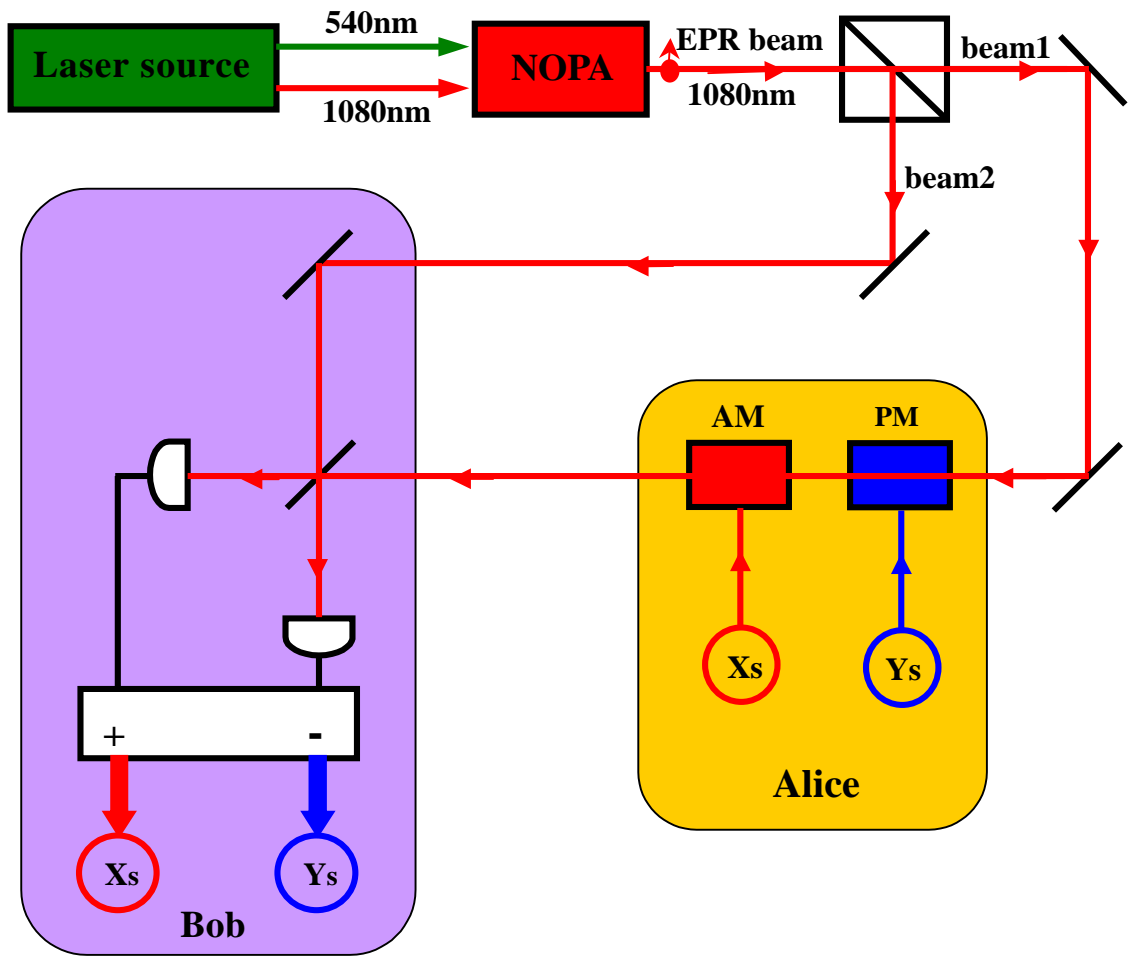

Figure.1 (X.Y. Li et al.) Schematic of the experimental apparatus for dense coding for Continuous variables. Two bits of classical information Xs and Ys are encoded on the amplitude and phase quadratures of a half of EPR beam (beam 1) at Alice, then are decoded by the other half of EPR beam (beam 2) at Bob.